\documentclass[twocolumn,aps,prl,showpacs,superscriptaddress]{revtex4}
\usepackage{graphicx}
\usepackage{amsmath,amssymb}
\usepackage{color}
\usepackage{multirow}

\newcommand{\ket}[1]{|#1\rangle}

\begin{document}


\title{Experimental observation of impossible-to-beat quantum advantage on a hybrid photonic system}


\author{Eleonora Nagali}
 \affiliation{Dipartimento di Fisica della ``Sapienza''
 Universit\`{a} di Roma, Roma 00185, Italy}
\author{Vincenzo D'Ambrosio}
 \affiliation{Dipartimento di Fisica della ``Sapienza''
 Universit\`{a} di Roma, Roma 00185, Italy}
\author{Fabio Sciarrino}
 \email{fabio.sciarrino@uniroma1.it}
 \homepage{http://quantumoptics.phys.uniroma1.it}
 \affiliation{Dipartimento di Fisica della ``Sapienza''
 Universit\`{a} di Roma, Roma 00185, Italy}
\author{Ad\'an Cabello}
 \email{adan@us.es}
 \affiliation{Departamento de F\'{\i}sica Aplicada II, Universidad de
 Sevilla, E-41012 Sevilla, Spain}
 \affiliation{Department of Physics, Stockholm University, S-10691
 Stockholm, Sweden}



\date{\today}



\begin{abstract}
Quantum resources outperform classical ones for certain communication and computational tasks. Remarkably, in some cases, the quantum advantage cannot be improved using hypothetical postquantum resources. A class of tasks with this property can be singled out using graph theory. Here we report
the experimental observation of an impossible-to-beat quantum advantage on a four-dimensional quantum system defined by the polarization and orbital angular momentum of a single photon. The results show pristine evidence of the quantum advantage and are compatible with the maximum advantage allowed using postquantum resources.
\end{abstract}




\maketitle


{\em Introduction.---}The search for properties singling out quantum mechanics from more general theories has recently attracted much attention \cite{Hardy01,PPKSWZ09,OW10,DB11,MM11,CDP11,BSBEW11,DMOS11}. In this framework, it is natural to address questions such as which is the simplest task in which quantum mechanics provides an advantage over classical theories and no hypothetical postquantum theory can do it better. The only requirement defining these postquantum theories is that they cannot assign a value larger than 1 to the sum of probabilities of mutually exclusive possibilities. Some recent results have shed light on this problem. Let us consider the class of tasks requiring one to maximize a sum $\Sigma$ of probabilities of propositions tested on a system (this class includes some communication complexity tasks \cite{BZPZ04} and all noncontextual \cite{Cabello08,KZGKGCBR09,ARBC09} and Bell inequalities). In Ref. \cite{CSW10} it is shown that the maximum of $\Sigma$ is given by $C(G)$, $Q(G)$, or $P(G)$, depending on whether classical, quantum, or general resources are used. This numbers are three properties of the graph $G$ in which vertices represent propositions and edges link exclusive propositions. The simplest task of this class in which there is a quantum advantage but no postquantum theory outperforms quantum mechanics corresponds to the simplest graph such that $C(G)<Q(G)=P(G)$, requiring a quantum system with the lowest possible dimensionality $\chi(G)$.

\begin{figure}[t]
\centerline{\includegraphics[width=4.4cm]{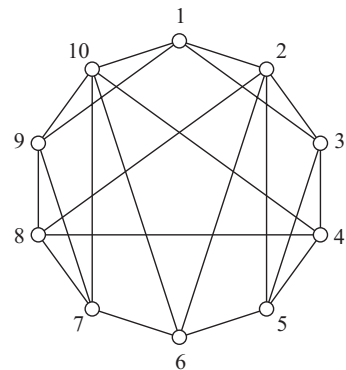}}
\caption{\label{c10}Graph representing the simplest task of the class defined in the main text with quantum but no postquantum advantage. Vertices represent propositions, edges link propositions that cannot be simultaneously true.}
\end{figure}

In this Letter we experimentally implement the simplest task with quantum but no postquantum advantage. For this purpose we exploit the properties of the graph of Fig. \ref{c10}, identified in \cite{CDLP11} as the simplest one with these properties, to perform an experiment in which quantum mechanics gives a larger $\Sigma$ than classical theories and no postquantum theory can do it better. Specifically, for the graph in Fig. \ref{c10}, $C(G)=3$ while $Q(G)=P(G)=3.5$ with $\chi(G)=4$. Experimentally we adopt a photonic hybrid system of dimension four, encoded in the polarization and a bidimensional subspace of the orbital angular momentum. The high fidelity and reliability of the present scheme allow us to achieve a close to theory measured value and a direct test of the exclusivity of the 10 involved orthogonal projectors.


There is a one-to-one correspondence between $C(G)$, $Q(G)$, and $P(G)$ and the classical, quantum, and general bounds for the following task: given an $n(G)$-vertex graph $G$, each player is asked to prepare a physical system and provide a list of $n(G)$ yes-no questions (or tests) $Q_i$ on this system, satisfying that questions corresponding to adjacent vertices in $G$ cannot both have the answer yes. The player who provides the preparation and questions with the highest probability of obtaining a yes answer when one question is picked at random wins.

If the questions refer to preexisting properties,that is, all the answers have a predefined value, the highest probability of obtaining a yes answer is $C(G)/n(G)$. For the graph in Fig. \ref{c10}, the sum of the probabilities of obtaining a yes answer is
\begin{equation}
 \label{inequality}
\Sigma =\sum_{i=1}^{10} P(Q_i=1) \le 3 = C(G),
\end{equation}
since at most 3 of the questions in Fig. \ref{c10} can be true. An optimal classical strategy to win is described in the Supplemental Material \cite{SM}.

However in quantum mechanics, preparing a four-level system in the state
\begin{equation}
 \langle \psi | = (0,0,0,1),
\end{equation}
and testing the propositions represented by the projectors $|v_i\rangle \langle v_i|$ over the following 10 (non-normalized) vectors $\langle v_i|$,
\begin{subequations}
 \label{states}
\begin{align}
&\langle v_1 | = (0,0,1,1),\\
&\langle v_2 | = (1,-1,1,-1),\\
&\langle v_3 | = (1,-1,-1,1),\\
&\langle v_4 | = (1,0,0,-1),\\
&\langle v_5 | = (1,1,1,1),\\
&\langle v_6 | = (0,1,0,-1),\\
&\langle v_7 | = (-1,1,1,1),\\
&\langle v_8 | = (1,0,0,1),\\
&\langle v_9 | = (1,1,1,-1),\\
&\langle v_{10} | = (1,1,-1,1),
\end{align}
\end{subequations}
the probability of obtaining a yes answer is $\frac{7}{20}=0.35$, which is the maximum using quantum resources [namely, $Q(G)/n(G)$], since for the graph in Fig. \ref{c10},
\begin{equation}
 Q(G)=\frac{7}{2}.
\end{equation}
which does not only go beyond the classical limit, but actually saturates the bound for any postquantum theory. The simplest way to grasp the previous bound is to notice that any other assignment of probabilities to the vertices of the graph in Fig. \ref{c10} either does not beat $7/2$ or is inconsistent with the requirement that the sum of probabilities of mutually adjacent vertices (i.e., those representing mutually exclusive propositions) cannot be larger than $1$. As explained in \cite{CSW10}, there is a one-to-one correspondence between the maximum of the sum of the probabilities and the so-called fractional packing number of the graph in which vertices represent propositions and edges exclusiveness. The fractional packing number of the graph in Fig. \ref{c10} is $7/2$. The remarkable property of the graph in Fig. \ref{c10} is that $Q(G)=P(G)$, so no postquantum theory can improve this performance. Unlike standard Bell tests where hypothetical postquantum theories can outperform quantum mechanics \cite{AGACVMC11}, here quantum mechanics reaches the maximum performance allowed by the laws of probability, as in this case there is no way to assign probabilities outperforming the quantum ones without violating that the sum of the probabilities of exclusive propositions cannot be higher than $1$. Indeed, what makes this experiment special is that it aims to the simplest scenario where the quantum probabilities exhibit this curious property.


{\em Experimental implementation.---}To experimentally verify the quantum predictions we require a four-dimensional system and the ability to project ququart states over all the states in Eqs. \eqref{states} with high fidelity and high reliability. These states are found to belong to all the five different mutually unbiased bases of a ququart \cite{Klap03,Plan07}. We encoded such higher-dimensional quantum states by exploiting two different degrees of freedom of the same photon. It has been recently demonstrated that ququart states can be efficiently generated by manipulating the polarization and orbital angular momentum (OAM) \cite{Naga10pra}. The orbital angular momentum of light is related to the photon's transverse-mode spatial structure \cite{Alle92} and can be exploited for implementing qudits encoded in a single photon state \cite{Moli07,Fran08,Aiel05}. The combined use of different degrees of freedom of a photon, such as OAM and spin, enables the implementation of entirely new quantum tasks \cite{Barr08}. Moreover, the implementation of a ququart state by exploiting both the polarization and a bidimensional subspace of orbital angular momentum with fixed OAM eigenvalue $|m|$, the so-called hybrid approach, does not require interferometric stability and is not affected by decoherence due to different Gouy phase for free propagation \cite{Naga09opt}.

Here we considered a bidimensional subset of the infinite-dimensional OAM space, denoted as $o_2$, spanned by states with OAM eigenvalue $m=\pm 2$ in units of $\hbar$. According to the nomenclature $\ket{\varphi,\phi}=\ket{\varphi}_{\pi}\ket{\phi}_{o_2}$, where $\ket{\cdot}_{\pi}$ and $\ket{\cdot}_{o_2}$ stand for the
photon quantum state ``kets'' in the polarization and OAM degrees of freedom, respectively, the logic ququart basis can be written as
\begin{equation}
\{\ket{H,+2},\ket{H,-2},\ket{V,+2},\ket{V,-2}\},
\end{equation}
where $H$ ($V$) refers to horizontal (vertical) polarization. According to these definitions, a generic ququart state expressed as $(a_1, a_2, a_3, a_4)$, as in \eqref{states}, can be experimentally implemented as
\begin{equation}
a_1\ket{H,+2}+a_2\ket{H,-2}+a_3\ket{V,+2}+a_4\ket{V,-2}.
\end{equation}
Analogously to \cite{Naga10pra}, the manipulation of the OAM degree of freedom has been achieved by adopting the $q$-plate device \cite{Naga09prl,Naga09opt}. On the polarization, the $q$ plate acts as a half-wave plate, while on the OAM it imposes a shift on the eigenvalue $m=\pm2q$, where $q$ is an integer or half-integer number determined by the (fixed) pattern of the optical axis of the device. In our experiments we adopted a $q$ plate with $q=1$, thus manipulating the OAM subspace $o_2=\{\ket{+2},\ket{-2}\}$. Interestingly, the ability of the $q$ plate to entangle and disentangle the OAM-polarization degrees of freedom can be exploited for the preparation as well as for the measurement of any ququart states.


\begin{figure*}[t]
\centerline{\includegraphics[scale=0.42]{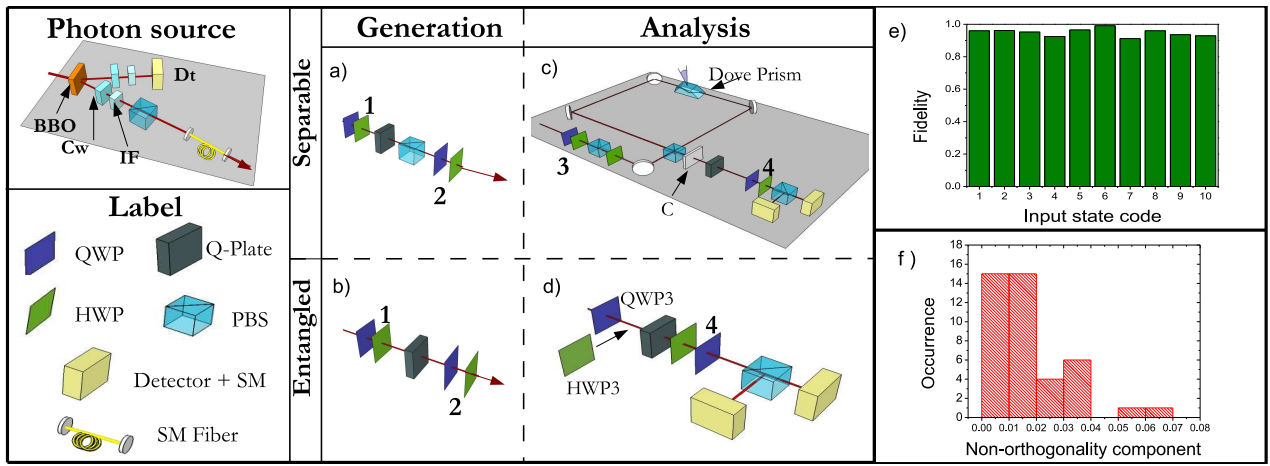}}
\caption{\label{setup} Experimental setup for the measurement of the probabilities $p_{i,j}$. In the upper left corner is represented the single photon source, based on spontaneous parametric down conversion in a non linear crystal (BBO) cut for type II SPDC generation of photon pairs and the compensation for the walk-off ($C_w$) followed by two interference filters (IF) with bandwidth $\Delta\lambda=3$ nm. See the Supplemental Material \cite{SM} for further details. The four schemes we used for the experiment are presented in the central part of the figure. Each state is prepared by one of the two setups of the Generation column: Setup (a) for separable states (quantum transferrer $\pi\rightarrow o_2$) and setup (b) for entangled ones [an ``entangler'' based on a $q$ plate, a
quarter-wave plate (QWP), and a half-wave plate (HWP)]. The Analysis column shows the setups for the projection on the desired state. Setup (c) for separable states, a deterministic transferrer $o_2\rightarrow \pi$ (here $C$ is a compensation stage), and setup (d) for entangled states, where a $q$ plate is needed to have a deterministic detection. (e) Experimental fidelities of generation and analysis for the ten states of the graph in Fig. \ref{c10}. (f) Experimental results of the exclusiveness test: occurrences of the non orthogonality component of the experimental projectors adopted for the measurements. All measured values of $p(i,j)$ and $p(j,i)$ are reported in the Supplemental Material \cite{SM}.}
\end{figure*}


Let us briefly give an example of how the $q$ plate works. By injecting a photon in the state $\ket{R}_{\pi}\ket{0}_o$ ($\ket{L}_{\pi}\ket{0}_o$), where $\ket{R}_{\pi}$ ($\ket{L}_{\pi}$) refers to the right (left) circular polarization, the output state reads $\ket{L}_{\pi}\ket{-2}_{o_2}$ ($\ket{R}_{\pi}\ket{+2}_{o_2}$). It follows from the latter relations that by injecting on a $q$ plate a photon with null OAM value and horizontal polarization, the state $(\ket{R,+2}+\ket{L,-2})/\sqrt{2}$ is generated, corresponding to a single photon entangled state between two different degrees of freedom.
In Table \ref{sets} we report the projections over the ten ququart states on the input state
$(0,0,0,1)=\ket{V,-2}$ needed to obtain the maximum possible violation in quantum mechanics.


\begin{table}[h]
{\small
\begin{tabular}{ccccc}
\hline\hline
 \multicolumn{1}{c}{} & \multicolumn{1}{c}{} & \multicolumn{1}{c}{} & \multicolumn{2}{c}{Probabilities}\\
State projection & Code & Type & {Theory} & {Experiment}\\
\hline
$(0,0,1,1)$ & 1 & $S$ & $1/2$ & $0.69\pm0.02$\\
 $(1,-1,1,-1)$ & 2 & $S$ & 1/4 & $0.160\pm0.007$\\
 $(1,-1,-1,1)$ & 3 & $S$ & 1/4 & $0.145\pm0.006$\\
 $(1,0,0,-1)$ & 4 & $E$ & 1/2 & $0.44\pm0.01$\\
 $(1,1,1,1)$ & 5 & $S$ & 1/4 & $0.33\pm0.01$\\
 $(0,1,0,-1)$ & 6 & $S$ & 1/2 & $0.49\pm0.01$\\
 $(-1,1,1,1)$ & 7 & $E$ & 1/4 & $0.160\pm0.007$\\
 $(1,0,0,1)$ & 8 & $E$ & 1/2 & $0.51\pm0.01$\\
 $(1,1,1,-1)$ & 9 & $E$ & 1/4 & $0.34\pm0.01$\\
 $(1,1,-1,1)$ & 10 & $E$ & 1/4 & $0.218\pm0.008$\\
&& Sum & 7/2 & $3.49\pm 0.03$\\
 \hline\hline
\end{tabular}
}
\caption{Theoretical predictions and experimental results for the probabilities of the different outcomes from measurements on state $(0,0,0,1)=\ket{V,-2}$. We associate to each projection a number used later to identify the state. In the column labeled Type we specify if the state is separable ($S$) or entangled ($E$).}
\label{sets}
\end{table}


The experimental setup adopted for such measurements is shown in Fig. \ref{setup}. A spontaneous parametric source (SPDC) generates heralded single photon states, sent through single mode (SM) fiber to setup (a) in order to encode the input state $(0,0,0,1)$, generated adopting a quantum transferrer $\pi\rightarrow o_2$. This tool allows us to transfer the information initially encoded in the polarization degree of freedom to the OAM, by exploiting the features of the $q$-plate device combined to a polarizing beam splitter (PBS) \cite{Naga09prl}. In particular, the input state has been generated by adopting the experimental setup in Fig. \ref{setup}(a), where the wave plates $1$ were oriented to generate right circular polarization, and the settings of wave plates $2$ for vertical polarization. Then, measurements have been carried out adopting the setups in Figs. \ref{setup}(c) and \ref{setup}(d), depending on whether the state on which the projection had to be carried out was separable or entangled. For the projection on separable states (denoted by $S$ in Table \ref{sets}), we adopted a deterministic transferrer $o_2\rightarrow\pi$ based on a Sagnac interferometer with a Dove's prism in one of its arms \cite{Damb11}. Thanks to this setup, any qubit encoded in a bidimensional subspace of OAM $\ket{\varphi}_{o_2}$ is transferred to the polarization with probability $p=1$. When the analysis on entangled states has to be carried out, it is possible to exploit the capability of the $q$ plate to disentangle the polarization to the OAM of a single photon. Indeed, for such projections we adopted a $q$ plate and a standard polarization analysis setup. The experimental results are reported in Table \ref{sets} and compared to the theoretical value of $3.5$. We observed a good agreement with the theoretical expectations, thus demonstrating the advantage of adopting quantum resources over classical ones.

As a second step, we provide the experimental verification of exclusiveness relations between the different states in \eqref{states}, that is, the fact that states connected by an edge cannot be simultaneously both true. We denote by a number from 1 to 10 the states involved in the experiment, and measured the probabilities $p(i,j)$ and $p(j,i)$, where $i,j=1,\ldots,10$. For the generation of ququart states belonging to entangled bases, we adopted the scheme reported in Fig. \ref{setup}(b). In Fig. \ref{setup}(e) we report the experimental values of probabilities $p(i,i)$, measured in order to ensure a high fidelity in the generation and reliability of all ququart states involved in the experiment. In particular, we observed an average fidelity of $F=(0.9492\pm0.0001)$. To verify that experimentally we implement orthogonal projectors, we measured the probabilities $p(i,j)$ and $p(j,i)$ with $i\neq j$. In Fig. \ref{setup}(f) we report the histogram of the occurrence of different values of probabilities, that quantify the nonorthogonality component of the experimental projectors. We observe a good agreement with the null value expected for orthogonal states. Error bars have been evaluated by considering the poissonian statistics of photon events.


{\em Discussion.---}The classical inequality \eqref{inequality} is valid under the assumption that the measured propositions satisfy the exclusiveness relations given by the graph in Fig. \ref{c10}. The results in Fig. \ref{setup}(f) show a very good agreement with the assumption. Even if the agreement between the experimental sum of probabilities is high, for some probabilities the deviations from the theoretical results are larger than the error bars. We attribute such discrepancy to the experimental implementation of the projectors, whose orientation with respect to the input state is slightly different from the expected one. Assuming that inequality \eqref{inequality} is only valid with probability $1-\epsilon$ and assuming that the worst case scenario, in which there are no links so the bound of the inequality is 10, occurs with probability $\epsilon$, to certify the quantum advantage it is enough that $3 (1-\epsilon) + 10 \epsilon < 3.49$; that is, $\epsilon<0.071$. The impossible-to-beat quantum advantage is certified by the fact that all our 42 experimental probabilities satisfy this condition and by the fact that the average value of $\epsilon$ is $0.016\pm0.001$. To our knowledge, this is the first time an experiment performing a task with quantum but no postquantum advantage \cite{AGACVMC11,ZBCYCP07,KSYZVCLJ11} has show results which demonstrate the quantum advantage and are compatible with the impossibility of a better performance.

In summary, in this Letter we reported the experimental implementation of the simplest impossible-to-beat quantum advantage by adopting a photonic system of dimension four. Such system has been implemented by exploiting the polarization and orbital angular momentum of single photons. We found a good agreement with theoretical expectation values. Moreover, we have experimentally verified all the exclusiveness relations between the states that correspond to the elements of the graph that models our system.


\begin{acknowledgments}
This work was supported by the MICINN Project No.~FIS2008-05596, the Wenner-Gren Foundation, FIRB Futuro in Ricerca-HYTEQ, and Project PHORBITECH of the Future and Emerging Technologies (FET) program within the Seventh Framework Programme for Research of the European Commission, under FET-Open Grant No.~255914.
\end{acknowledgments}


\end{document}